\documentclass[a4paper]{llncs}

\usepackage[utf8x]{inputenc}

\usepackage{amsmath}

\usepackage{amsthm}
\usepackage{amssymb}
\usepackage{breqn}
\usepackage{framed}
\usepackage{array}
\usepackage{xfrac}
\usepackage{tikz}
\usepackage{tkz-berge}
\usepackage{enumerate}
\usepackage{bookmark}
\usepackage{hyperref}
\hypersetup{
  pdftitle = {Graph Collision},
  pdfauthor = {},
  pdfkeywords = {},
  pdfstartview=FitH,
  unicode=true
}

\newcommand{\ket}[1]{\left| #1 \right\rangle}
\newcommand{\bra}[1]{\left\langle #1 \right|}
\newcommand{\braket}[2]{\left\langle #1 | #2 \right\rangle}

\begin{document}

\title{Parameterized Quantum Query Complexity of Graph Collision
\thanks{%
This work has been supported by the European Social Fund within the project ``Support for Doctoral Studies at University of Latvia'' and by the FP7 FET-Open project QCS.
}
}
\author{\texorpdfstring{Andris Ambainis \and Kaspars Balodis\and Jānis Iraids \and Raitis Ozols \and Juris Smotrovs}{Andris Ambainis, Kaspars Balodis, Jānis Iraids, Raitis Ozols, Juris Smotrovs}}

\institute{University of Latvia, Raiņa bulvāris 19, Riga, LV-1586, Latvia}

\maketitle

\begin{abstract}
We present three new quantum algorithms in the quantum query model for \textsc{graph-collision} problem:
\begin{itemize}
\item an algorithm based on tree decomposition that uses $O\left(\sqrt{n}t^{\sfrac{1}{6}}\right)$ queries where $t$ is the treewidth of the graph;
\item an algorithm constructed on a span program that improves a result by Gavinsky and Ito. The algorithm uses $O(\sqrt{n}+\sqrt{\alpha^{**}})$ queries, where $\alpha^{**}(G)$ is a graph parameter defined by
\[\alpha^{**}(G):=\min_{VC\text{ -- vertex cover of }G}{\max_{\substack{I\subseteq VC\\I\text{ -- independent set}}}{\sum_{v\in I}{\deg{v}}}};\]
\item an algorithm for a subclass of circulant graphs that uses $O(\sqrt{n})$ queries.
\end{itemize}
We also present an example of a possibly difficult graph $G$ for which all the known
graphs fail to solve graph collision in $O(\sqrt{n} \log^c n)$ queries.
\end{abstract}

\section{Introduction}
\textsc{graph-collision} (or $COL(G)$) is a decision problem in which we are given an undirected graph $G$ and black-box access to boolean variables $\{x_v \mid v\in V(G)\}$, where $V(G)$ is the set of vertices of $G$, $|V(G)|=n$. $COL(G)$ asks whether there are two vertices $v$ and $v'$ connected by an edge in $G$, such that $x_v=x_{v'}=1$.

The graph collision problem was introduced in \cite{MSS07} and an algorithm for graph collision was used as a subroutine for an algorithm for triangle-finding. The quantum query complexity of the subroutine is $O(n^{\sfrac{2}{3}})$ and this is still the best known upper bound for graph collision for unrestricted graphs $G$. On the other hand, the best known lower bound is the trivial $\Omega(\sqrt{n})$ which follows by embedding Grover's search problem (in a star graph, for instance). 

There is a number of algorithms for special cases of graph collision.
For many of them, their complexities depend on additional parameters of the graph.
\begin{center}
  \begin{tabular}{|p{2.5cm}|p{6cm}|p{3cm}|}
    \hline
    Complexity & Parameters & Discoverer \\ \hline \hline
    $\tilde{O}(\sqrt{n}+\sqrt{\ell})$ & $\ell$ - the number of non-edges in the graph & Jeffery et al. \cite{JKM12} \\ \hline
    $O(\sqrt{n}\alpha^{\sfrac{1}{6}})$ & $\alpha$ - the size of largest independent set of $G$ & Belovs \cite{Bel12a} \\ \hline
    $O\left(\sqrt{n}+\sqrt{\alpha^*}\right)$ & $\alpha^*$ - the maximum total degree of any independent set of $G$ & Gavinsky et al. \cite{GI12} \\ \hline
    $\tilde{O}(\sqrt{n})$ & random graphs with fixed probability of each edge independently of the others & Gavinsky et al. \cite{GI12} \\ \hline
    $O(\sqrt{m})$ & $m$ - the number of edges in $G$ & trivial \\ \hline
  \end{tabular}
\end{center}

In this paper we present 3 more parameterized algorithms:
\begin{center}
  \begin{tabular}{|p{3cm}|p{8cm}|}
    \hline
    Complexity & Parameters  \\ \hline \hline
    $O(\sqrt{n}t^{\sfrac{1}{6}})$ & $t$ - the treewidth of graph $G$ \\ \hline
    $O(\sqrt{n}+\sqrt{\alpha^{**}})$ & \[\alpha^{**}=\min_{VC\text{ -- vertex cover of }G}{\max_{\substack{I\subseteq VC\\I\text{ -- independent set}}}{\sum_{v\in I}{\deg{v}}}}\] \\ \hline
    $O(\sqrt{n})$ & $G$ belongs to a certain subclass of circulant graphs (see below) \\ \hline
  \end{tabular}
\end{center}

\section{Graph collision based on treewidth}

In \cite{Bel12a} Belovs showed a learning graph based algorithm for graph collision with $O(\sqrt{n}\alpha^{\sfrac{1}{6}})$ queries where $\alpha$ is the size of the largest independent set in the graph.
As $\alpha$ can be at most $n$ the maximum number of queries needed for any graph is at most $O(n^{\sfrac{2}{3}})$.
We show a quantum algorithm that uses $O(\sqrt{n}t^{\sfrac{1}{6}})$ queries where $t$ is the treewidth of the graph.

\begin{theorem}
Graph collision on graph $G$ on $n$ vertices can be solved with a bounded error quantum algorithm with $O(\sqrt{n}t^{\sfrac{1}{6}})$ queries where $t$ is the treewidth of the graph.
\end{theorem}

The notion of treewidth was (re)introduced by Robertson and Seymour \cite{RS86treewidth} in their series of graph minors.

\begin{definition}
A tree decomposition of a graph $G=(V,E)$ is a pair $(\{X_i \mid i \in I \}, T=(I,F))$ with $\{X_i \mid i \in I \}$ a family of subsets of $V$ (called bags), one for each node of $T$, and $T$ a tree such that
\begin{itemize}
\item $\bigcup_{i\in I}{X_i=V}$
\item for all edges $(v,w) \in E$, there exists an $i \in I$ with $v \in X_i$ and $w \in X_i$
\item for all $i,j,k\in I$: if $j$ is on the path from $i$ to $k$ in $T$, then $X_i \cap X_k \subseteq X_j$.
\end{itemize}
The treewidth of a tree decomposition $(\{X_i \mid i \in I\}, T=(I,F))$ is 
\[\max_{i\in I}{|X_i|-1}.\]
The treewidth of a graph $G$ is the minimum treewidth over all possible tree decompositions of $G$.

We say a tree decomposition $(X,T)$ of treewidth $k$ is smooth, if for all $i\in I$: $|X_i|=k+1$, and for all $(i,j)\in F:|X_i\cap X_j|=k$. Any tree decomposition can be transformed to a smooth tree decomposition with the same treewidth. Moreover for a smooth tree decomposition: $|I|=|V|-k$. \cite{B96treewidth}
\end{definition}

The idea of the algorithm is to use Grover search over the vertices of a bit modified decomposition of the graph $G$.
The algorithm has preprocessing part where the vertices of the graph are divided into $O(\frac{n}{t})$ subsets of size $O(t)$ in such a way that if there is a collision in $G$ then there is also a collision in at least one of the induced subgraphs of the subsets.
The algorithm then uses Grover search \cite{Gro96} to find a subset whose induced subgraph has a collision. As a subroutine a graph collision algorithm that uses $O(n^{\sfrac{2}{3}})$ queries for a graph on $n$ vertices for is used (for example the algorithm by Belovs \cite{Bel12a} can be used).

\begin{lemma}\label{decompose}
For a graph $G=(V,E)$ with treewidth $t$ there exists a decomposition into subsets of vertices $W=(W_i)$, $W_i \subseteq V$ such that $|W|=\left\lceil \frac{2n-2t-1}{t} \right\rceil$, for all $i$: $|W_i|\leq 2t+1$ and for each edge $(v,w) \in E$: $\exists i: v,w \in W_i$.
\end{lemma}
\begin{proof}
Let $(\{X_i \mid i \in I\}, T=(I,F))$ be a smooth tree decomposition of $G$ with treewidth $t$.

The tree $T$ can be transformed to a sequence $U = (u_i)$, $u_i \in I$ of length $2n-2t-1$ of vertices of $T$ with $(u_i, u_{i+1}) \in F$ by drawing $T$ on a plane and traversing it like a maze by following the left wall (that is for the next edge to traverse choosing the one which makes the smallest angle clockwise with the last traversed edge) and finishing when every edge has been traversed exactly twice.
The sequence of the traversed vertices is $U$.

Next, slice the sequence $U$ into parts of length $t$ and merge the corresponding vertices of $G$ into subsets $W_i = \bigcup_{t\cdot (i-1) \leq j < t \cdot i}{X_{u_j}}$.

As each $W_i$ consists of vertices from $t$ incident tree decomposition bags and two incident bags differ by one element, the size of each $|W_i|$ is at most $2t+1$.

As for each edge $(u,v)\in E$ there exists a bag that contains both $u$ and $v$ and the subsets $W_i$ are unions of all bags then for each edge $(u,v)\in E$ there exists a subset $W_i$ that contains both $u$ and $v$.
\qed
\end{proof}

The algorithm:
\begin{enumerate}
\item (Preprocessing) Decompose the graph $G$ into subsets of vertices $W_i$ as in Lemma \ref{decompose}.
\item Perform a Grover search for $i$ such that $G[W_i]$ -- the subgraph of $G$ induced by $W_i$ -- contains a collision. As a subroutine use an algorithm that detects a collision in $G[W_i]$ with $|W_i|^{\sfrac{2}{3}}$ queries.
\end{enumerate}

\subsection*{Query complexity}
The preprocessing part depends only on the graph $G$ instead of the values of the associated variables therefore it does not use any queries.

The Grover search calls the subroutine $O(\sqrt{m})$ times where $m$ is the number of subsets. Therefore the total number of queries for the algorithm is at most $O( \sqrt{\left\lceil \frac{2n-2t-1}{t} \right\rceil} \cdot (2t+1)^{\sfrac{2}{3}}) = O(\sqrt{\frac{n}{t}}\cdot t^{2/3}) = O(\sqrt{n}t^{1/6})$.

\section{A span program for graph collision}

Span programs is a model of computation introduced in \cite{KW93} by Karchmer and Widgerson in 1993. Span programs are used to evaluate a boolean function. In 2009 Reichardt showed \cite{Rei09,Rei11} that there is a strong connection between a complexity measure of span programs -- span program's witness size -- and quantum query complexity. Since then span programs have been of great interest in quantum computing. Belovs used span programs to construct quantum algorithms from learning graphs \cite{Bel12,Bel12a}.

Recently, Gavinsky and Ito have used constructed span programs for graph collision \cite{GI12}. Denote by $\alpha^*(G)$ the maximum total degree of any independent set of $G$.

\begin{theorem}[Gavinsky, Ito\cite{GI12}]
\[Q(COL(G)) = O\left(\sqrt{|V(G)|}+\sqrt{\alpha^*(G)}\right)\]
\end{theorem}

We improve their result by constructing a simpler span program that works in a more general case, and we give an example for which our algorithm performs better. In addition, we do not require using quantum counting as a subroutine.

\subsection{Overview of span programs}
\begin{definition}
A \emph{span program} $P$ is a tuple $P=(H, \ket{t}, V)$, where $H$ is a finite-dimensional Hilbert space, $\ket{t}\in H$ is called the target vector, and $V=\{V_{i,b}|i \in [n], b \in \{0,1\}\}$, where each $V_{i,b} \subseteq H$ is a finite set of vectors.

Denote by $V(x) = \bigcup{\{V_{i,b}|i\in [n], x_i=b\}}$. The span program is said to \emph{compute} function $f:D \rightarrow \{0,1\}$, where the domain $D \subseteq \{0,1\}^n$, if for all $x\in D$, 
   \[f(x)=1 \iff \ket{t} \in \operatorname{span}(V(x)).\]
\end{definition}

It may be helpful to view $V$ as a set of vectors with labels that put a constraint on at most one input bit. Note that vectors with no constraints can be allowed by including the same vector in both $V_{1,0}$ and $V_{1,1}$. A vector can then be used depending on whether input $x$ satisfies vector's constraint.

\begin{definition}

\begin{enumerate}[(1)]
\item A \emph{positive witness} for $x\in f^{-1}(1)$ is a vector $w=(w_v),v\in V(x)$, such that $\ket{t}=\sum_{v\in V(x)}{w_v v}$. The \emph{positive witness size} is 
\[wsize_1(P):=\max_{x\in f^{-1}(1)}{\min_{w:\text{witness of }x}{\|w\|^2}}.\]

\item A \emph{negative witness} for $x\in f^{-1}(0)$ is a vector $w\in H$, such that $\braket{t}{w}=1$ and for all $v\in V(x)$: $\braket{v}{w}=0$. The \emph{negative witness size} is \[wsize_0(P):=\max_{x\in f^{-1}(0)}{\min_{w:\text{witness of }x}{\sum_{v\in V}{\braket{v}{w}}}}.\]

\item The \emph{witness size of a program} $P$ is
\[wsize(P):=\sqrt{wsize_0(P)\cdot wsize_1(P)}.\]

\item The \emph{witness size of a function} $f$ denoted by $wsize(f)$ is the minimum witness size of a span program that computes $f$.
\end{enumerate}
\end{definition}

\begin{theorem}[\cite{Rei09,Rei11}]
\label{thm:span}
$Q(f)$ and $wsize(f)$ coincide up to a constant factor. That is, there exists a constant $c > 1$ which does not depend on $n$ or $f$ such that $\frac{1}{c}wsize(f) \leq Q(f) \leq c \cdot wsize(f)$.
\end{theorem}

\subsection{Span program for graph collision}

Denote by $N(v)$ be the set of neighbours of vertex $v$ in graph $G$. Let $VC\subseteq V(G)$ be a vertex cover of $G$. The span program is as follows:
\begin{framed}
\begin{center}Span program $P$ for graph collision
\end{center}
\begin{itemize}
\item $H$ is a $(|VC|+1)$ dimensional vector space with basis vectors $\{\ket{0}\}\cup\{\ket{v} | v \in VC\}$.\par
\item The target vector will be $\ket{0}$.\par
\item For all $v \in VC$, such that $x_v=1$, make available the vector $\ket{0}-\ket{v}$.\par
\item For all $v \in VC$, for all $v' \in N(v)$, such that $x_{v'}=1$, make available the vector $\ket{v}$.\par
\end{itemize}
\end{framed}

It is easy to see that $P$ indeed computes $COL(G)$. It remains to calculate the witness size of $P$. If there is a collision in the graph, it must occur between a vertex from the vertex cover VC and some other vertex. Therefore a positive witness on edge $(v,v')$ of collision with $v\in VC$ is just $1\cdot (\ket{0}-\ket{v})+1\cdot \ket{v}=\ket{0}$. And so
\[wsize_0(P) \leq 1^2+1^2\leq 2.\]
As the negative witness $w$ we will pick the vector 
\[w=\ket{0}+\sum_{\substack{v\in VC\\x_v=1}}{\ket{v}}.\]
First we check that it indeed is a negative witness and calculate all the scalar products along the way:
\begin{enumerate}
\item $\braket{0}{w}=1$.
\item The available vectors of the form $\ket{0}-\ket{v}$ are exactly the vectors for which $w$ contains $\ket{v}$, therefore $\braket{\bra{0}-\bra{v}}{w}=0$. For the vectors that are not available, the scalar product is 1, and there are at most $|VC|\leq |V(G)|$ such vectors.
\item The available vectors of the form $\ket{v}$ are only available if $x_v=0$, because otherwise there would be a collision between $v$ and some neighbour of $v$. Therefore for available $\ket{v}$:$\braket{v}{w}=0$. On the other hand, if the vector is not available, scalar product with $w$ is 1. For any two vectors $\ket{v}\neq \ket{v'}$, if $\braket{v}{w}=\braket{v'}{w}=1$, $x_v=1$ and $x_v'=1$, so there can be no edge $(v,v')$. Therefore the vectors that have scalar product $1$ with the negative witness must correspond to vertices of an independent set. Consequently, the total number of vectors that have $\braket{v}{w}=1$ is at most
\[D := \max_{\substack{I\subseteq VC\\I\text{ -- independent set}}}{\sum_{v\in I}{\deg{v}}}.\]
\end{enumerate}
Finally, we choose the vertex cover $VC$ for constructing $P$ in such a way to minimize $D$. Denote the resulting expression:
\[\alpha^{**}(G):=\min_{VC\text{ -- vertex cover of }G}{\max_{\substack{I\subseteq VC\\I\text{ -- independent set}}}{\sum_{v\in I}{\deg{v}}}}\]
Here we point out that $\alpha^{**}(G)\leq \alpha^*(G)$, since we get $\alpha^*(G)$ as a subcase of taking all vertices of $G$ as the vertex cover when minimizing over all vertex covers.
Our program winds up with the negative witness size of 
\[wsize_0(P) \leq |V(G)| + \alpha^{**}(G).\]

The witness size of the graph collision can be upper bounded by the witness size of our program:
\[\begin{split}wsize(COL(G)) \leq wsize(P) = \sqrt{wsize_0(P)\cdot wsize_1(P)} =\\ 
= O\left(\sqrt{|V(G)|+\alpha^{**}(G)}\right) = O\left(\sqrt{|V(G)|}+\sqrt{\alpha^{**}(G)}\right).
\end{split}
\]

From Theorem \ref{thm:span}, we arrive at
\begin{theorem}
\[Q(COL(G))= O\left(\sqrt{|V(G)|}+\sqrt{\alpha^{**}(G)}\right).\]
\end{theorem}

\subsection{Improvement for explicitly specified graphs}

\begin{figure}
  \begin{center}
    \includegraphics[scale=0.8]{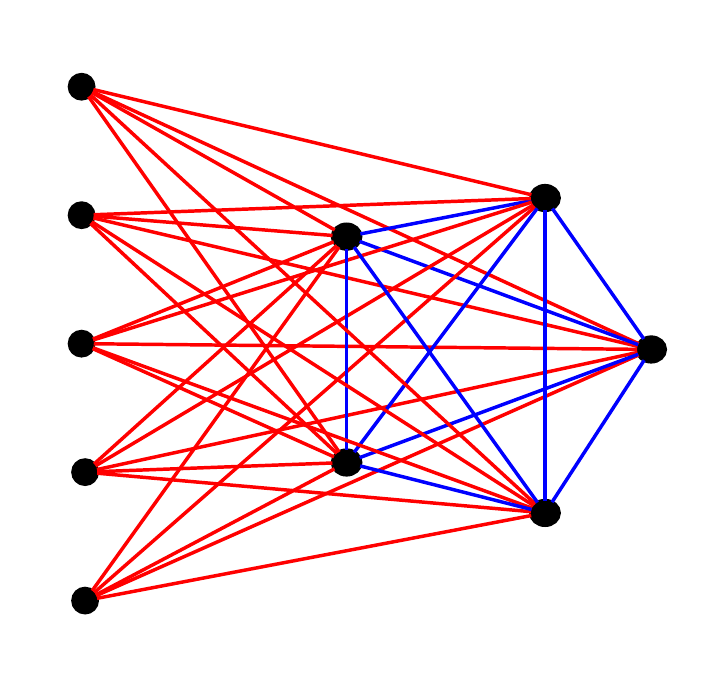}
  \end{center}
  \caption{The join of $\overline{K_5}$ and $K_5$}
  \label{fig:kk}
\end{figure}

In this section we give an example of a graph class for which algorithm performs better than the existing algorithms and in particular Gavinsky's and Ito's algorithm. The graph is a join of $\overline{K_n}$ and $K_n$ (see Figure \ref{fig:kk}). For this class of graphs:
\[\alpha^*(G)=n^2;\]
\[\alpha^{**}(G) \leq 2n.\]
Therefore our algorithm takes on the order of $O(\sqrt{n})$ queries, while their algorithm uses $O(n)$ queries.

\section{Quantum algorithm for a subclass of circulant graphs}

\subsection{Quantum algorithm}

In this section we desribe yet another quantum algorithm for a class of graphs that existing algorithms fail to solve efficiently. This algorithm uses Grover's algorithm for searching \cite{Gro96} and quantum algorithm for finding the minimum \cite{DH96}.

\begin{definition}
For any positive integers $n,a,b$ such that $a\leq b \leq \frac{n}{2}$, the graph $CI(n,a,b)$ is a circulant graph of $n$ vertices labeled with integers $0,1,\ldots, n-1$; an edge between two vertices $k,l$ ($k<l$) is present if and only if 
\[\min{\{l-k,n-(l-k)\}}\in [a,b].\]
\end{definition}

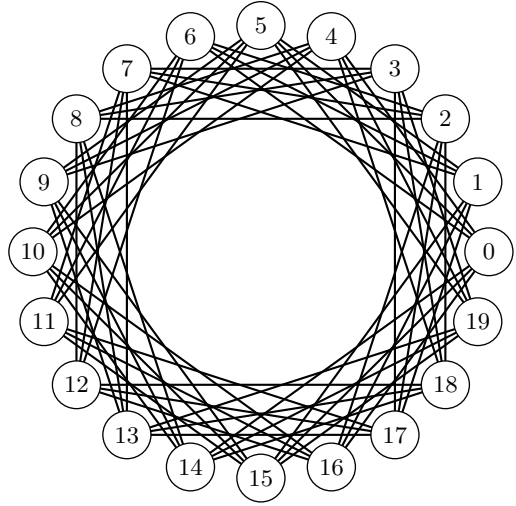
\begin{figure}
\begin{center}
\begin{tikzpicture}
  \grCirculant[RA=3,prefix=]{20}{4,5,6}
\end{tikzpicture}
\end{center}
\caption{The graph $CI(20,4,6)$}
\label{fig:ci2046}
\end{figure}

If we draw the vertices on a circle in an increasing order (except $n-1$ and $0$), there is an edge between vertices $k$ and $l$ if there are from $a-1$ to $b-1$ vertices between $k$ and $l$ on the circle. See, for example, Figure \ref{fig:ci2046}. 

The algorithm for $COL(CI(n,a,b))$ is the following: subdivide the vertices into consecutive chunks of size $b-a$:$V_1, V_2, \ldots, V_{\frac{n}{b-a}}$. Suppose we had an algorithm $SUB$, that given $V_i$ calculates whether there is a collision between some vertex $v \in V_i$ and some other vertex (we can assume that it is closer in the counterclockwise order from $v$) in $O(\sqrt{b-a})$ queries. Then using Grover's search over $SUB(V_i)$, we could in time $O\left(\sqrt{\frac{n}{b-a}}\sqrt{b-a}\right)=O(\sqrt{n})$ determine $COL(CI(n,a,b))$. The remainder of this section describes the algorithm $SUB$.

\begin{figure}
\begin{center}
\begin{tikzpicture}
\draw (0,0) circle (100pt);
\draw[very thick] (60:100pt) ++(0,0) arc (60:100:100pt);
\draw (60:95pt) -- (60:105pt);
\draw (60:110pt) node [rotate=60,anchor=west] {$k$};
\draw (100:95pt) -- (100:105pt);
\draw (100:110pt) node [rotate=110,anchor=west]{$k+(b-a)$};
\fill (70:100pt) circle (2pt);
\draw (70:110pt) node [rotate=70,anchor=west] {rightmost 1};
\fill (95:100pt) circle (2pt);
\draw (95:110pt) node [rotate=95,anchor=west] {leftmost 1};
\filldraw[fill=blue!80,draw=blue!80,opacity=0.5] (185:100pt) -- (70:100pt) arc (70:30:100pt) -- cycle;

\filldraw[fill=blue!80,draw=blue!80,opacity=0.5] (250:100pt) -- (95:100pt) arc (95:135:100pt) -- cycle;

\filldraw[fill=blue!20,draw=blue!80,opacity=0.3,style=dashed] (215:100pt) -- (60:100pt) arc (60:100:100pt) -- cycle;

\draw[very thick,draw=blue!100] (185:100pt) ++(0,0) arc (185:250:100pt);
\draw (185:95pt) -- (185:105pt);
\draw (185:110pt) node [rotate=5,anchor=east] {$u$};
\draw (250:95pt) -- (250:105pt);
\draw (250:110pt) node [rotate=70,anchor=east] {$v$};

\end{tikzpicture}
\end{center}
\caption{Illustration of subroutine SUB}
\label{fig:substep}
\end{figure}
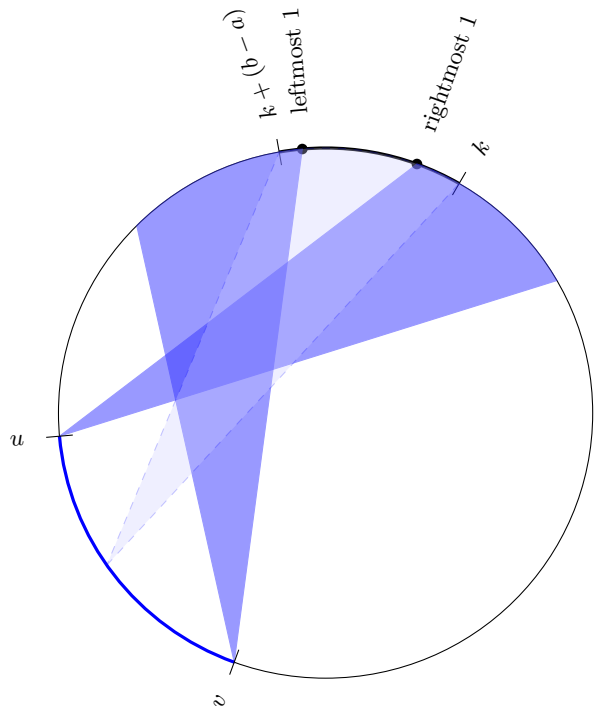

Consider the task $SUB$ is up to. Assume that the chunk starts at vertex $k$ and ends at vertex $k+(b-a)$. Use algorithm for $MIN(x_k, x_{k+1}, \ldots, x_{k+b-a})$ to find the smallest $r$, such that $x_r=1$. It takes $O(\sqrt{b-a})$ queries \cite{DH96}. If we find no $x_r=1$ in the chunk, we are done and return that there are no collisions on the vertices in the chunk. Otherwise continue and similarly find the largest $l$ in the chunk, such that $x_l=1$. Now assume that vertex $v$ is at distance $b$ from $l$ and $u$ is at distance $a$ from $r$ (see Figure \ref{fig:substep}). No vertex $s<u$ or $s>v$ is connected with any vertex from $r$ to $l$. On the other hand, if there is $s\in [u,v]$ with $x_s=1$ then it must be connected with at least one of $l$ and $r$. So, we have 

\begin{proposition}For some vertex $s\in [u,v]$:$x_s=1$, if and only if there is a collision for some vertex in the chunk.
\end{proposition}
There are at most $3(b-a)$ vertices between $u$ and $v$. Determining whether there is a vertex $s\in [u,v]$ with $x_s=1$ can be done by Grover's search in $O(\sqrt{b-a})$ queries.

\begin{theorem}
For graphs $CI(n,a,b)$:
\[Q(COL(CI(n,a,b))) = O(\sqrt{n}).\]
\end{theorem}

\section{Example of a potentially ``difficult'' graph}

For all concrete graphs that we know, one of the existing algorithms solve graph
collision with $O(\sqrt{n})$ (or $O(\sqrt{n \log^c n})$) queries. 
In this section, we present, possibly, 
the first example of an explicit  graph for which none of 
the existing quantum algorithms finds graph collision with substantially less
than $O(n^{2/3})$ queries.

\begin{definition}
A graph $CS(n)$ is graph with $n$ vertices labeled by integers from $0$ to $n-1$; there is an edge between vertices $k$ and $l$ if $|k-l|$ is a perfect square not exceeding $\frac{n}{2}$.
\end{definition}

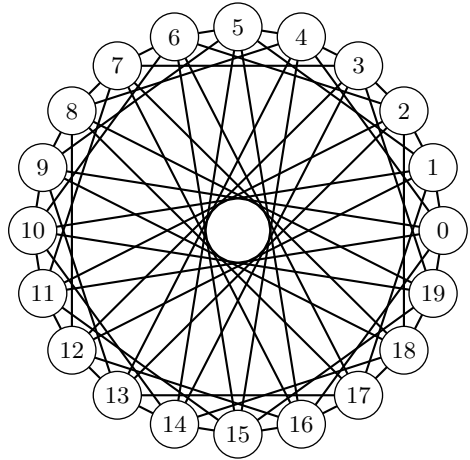
\begin{figure}
\begin{center}
\begin{tikzpicture}
  \grCirculant[RA=2.7,prefix=]{20}{1,4,9}
\end{tikzpicture}
\end{center}
\caption{The graph CS(20)}
\label{fig:cs20}
\end{figure}

For an example, see Figure \ref{fig:cs20}, that shows $CS(20)$ with edge ``lengths'' $1^2$, $2^2$ and $3^2$. The graph $CS(n)$ has $\Theta(n\cdot \sqrt{n})=\Theta(n^{\sfrac{3}{2}})$ edges.

\begin{figure}
\centering
\resizebox{0.55\textwidth}{!}{\input{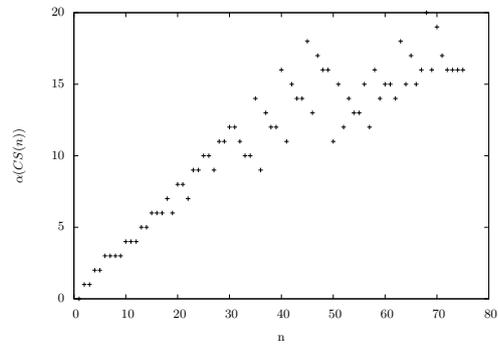}}
\caption{The independence number $\alpha(CS(n))$ for $n=1\ldots 75$}
\label{fig:sqind}
\end{figure}

Let $\alpha$ be the size of the largest independent set in $CS(n)$ (see Figure \ref{fig:sqind}).
Then, the best known bounds on $\alpha$ are:
\begin{itemize}
\item
$\alpha = \Omega(n^{0.7334...})$ \cite{BG08};
\item
$\alpha = O(\frac{n}{\log^c n})$ \cite{PSS88}.
\end{itemize}
Thus, the $O(\sqrt{n}\alpha^{1/6})$ bound on the number of queries 
in the algorithm of Belovs \cite{Bel12a} is between 
$O(n^{0.6222...})$ and $O(\frac{n^{2/3}}{\log^c n})$.

For the algorithm of Gavinsky et al. \cite{GI12} (or its improvement in this paper), we have 
$\alpha^*=\Theta(\alpha \sqrt{n})$, because every vertex in $CS(n)$ has outdegree $\Theta(\sqrt{n})$. Hence, the $O(\sqrt{n}+\sqrt{\alpha^*})$ bound on the
number of queries \cite{GI12} is between
$O(\sqrt{\sqrt{n} \cdot n^{0.7334...}})=O(n^{0.6167...})$
and $O(\frac{n^{3/4}}{\log^c n})$.

\section{Conclusion}

We are most optimistic about the approach of span programs. The main reason for our optimism is that the span program was simple. If a better span program exists, we should be able to find it using only slightly more refined methods.

The quantum query complexity of graph collision is an interesting open problem. While coming up with more efficient algorithms than $O(n^{\sfrac{2}{3}})$ for specific classes of graphs seems quite easy -- and indeed it is perhaps not surprising given that there exists $O(\sqrt{n})$ algorithm for most graphs -- no one has managed to construct an $O(\sqrt{n})$ algorithm for the general case.

\bibliographystyle{splncs03}

\phantomsection
\addcontentsline{toc}{chapter}{References}
\bibliography{../../bibliography/quantum,tw}

\begin{thebibliography}{10}
\providecommand{\url}[1]{\texttt{#1}}
\providecommand{\urlprefix}{URL }

\bibitem{BG08}
Beigel, R., Gasarch, W.: Square-difference-free sets of size omega
  (n\^{}$\{$0.7334...$\}$). arXiv preprint arXiv:0804.4892  (2008),
  \url{http://arxiv.org/abs/0804.4892}

\bibitem{Bel12a}
Belovs, A.: Learning-graph-based quantum algorithm for k-distinctness. In: IEEE
  53rd Annual Symposium on Foundations of Computer Science (FOCS). pp.
  207--216. IEEE (2012),
  \url{http://ieeexplore.ieee.org/xpls/abs_all.jsp?arnumber=6375298}

\bibitem{Bel12}
Belovs, A.: Span programs for functions with constant-sized 1-certificates. In:
  Proceedings of the 44th symposium on Theory of Computing. pp. 77--84. ACM
  (2012), \url{http://dl.acm.org/citation.cfm?id=2213985}

\bibitem{B96treewidth}
Bodlaender, H.L.: A linear-time algorithm for finding tree-decompositions of
  small treewidth. SIAM Journal on computing  25(6),  1305--1317 (1996)

\bibitem{DH96}
Dürr, C., Høyer, P.: A quantum algorithm for finding the minimum. arXiv
  preprint quant-ph/9607014  (1996),
  \url{http://arxiv.org/abs/quant-ph/9607014}

\bibitem{GI12}
Gavinsky, D., Ito, T.: A quantum query algorithm for the graph collision
  problem. arXiv preprint arXiv:1204.1527  (2012),
  \url{http://arxiv.org/abs/1204.1527}

\bibitem{Gro96}
Grover, L.K.: A fast quantum mechanical algorithm for database search. In:
  Proceedings of the twenty-eighth annual ACM symposium on Theory of computing.
  pp. 212--219. ACM (1996), \url{http://dl.acm.org/citation.cfm?id=237866}

\bibitem{JKM12}
Jeffery, S., Kothari, R., Magniez, F.: Improving quantum query complexity of
  boolean matrix multiplication using graph collision. In: Automata, Languages,
  and Programming, pp. 522--532. Springer (2012),
  \url{http://link.springer.com/chapter/10.1007/978-3-642-31594-7_44}

\bibitem{KW93}
Karchmer, M., Wigderson, A.: On span programs. In: Proceedings of the Eighth
  Annual Structure in Complexity Theory Conference. pp. 102--111. IEEE (1993),
  \url{http://ieeexplore.ieee.org/xpls/abs_all.jsp?arnumber=336536}

\bibitem{MSS07}
Magniez, F., Santha, M., Szegedy, M.: Quantum algorithms for the triangle
  problem. SIAM Journal on Computing  37(2),  413--424 (2007),
  \url{http://epubs.siam.org/doi/abs/10.1137/050643684}

\bibitem{PSS88}
Pintz, J., Steiger, W.L., Szemer{\'e}di, E.: On sets of natural numbers whose
  difference set contains no squares. Journal of the London Mathematical
  Society  2(2),  219--231 (1988),
  \url{http://jlms.oxfordjournals.org/content/s2-37/2/219.short}

\bibitem{Rei09}
Reichardt, B.W.: Span programs and quantum query complexity: The general
  adversary bound is nearly tight for every boolean function. In: 50th Annual
  IEEE Symposium on Foundations of Computer Science (FOCS). pp. 544--551. IEEE
  (2009), \url{http://ieeexplore.ieee.org/xpls/abs_all.jsp?arnumber=5438598}

\bibitem{Rei11}
Reichardt, B.W.: Reflections for quantum query algorithms. In: Proceedings of
  the Twenty-Second Annual ACM-SIAM Symposium on Discrete Algorithms. pp.
  560--569. SIAM (2011), \url{http://dl.acm.org/citation.cfm?id=2133080}

\bibitem{RS86treewidth}
Robertson, N., Seymour, P.D.: Graph minors. ii. algorithmic aspects of
  tree-width. Journal of algorithms  7(3),  309--322 (1986)

\end{thebibliography}

\end{document}